# Frequency Range Boosted Magnetometry Beyond the Spin Coherence Limit via Compressive Sensing

Ruiqi Wang[1], Peiyu Yang[1,3], Guzhi Bao[1,2,3*], and Weiping Zhang[1,2,3,4,5*]

[1] School of Physics and Astronomy, Shanghai Jiao Tong University,

Shanghai, 200240, China.

[2] Tsung-Dao Lee Institute, Shanghai Jiao Tong University, Shanghai, 200240, China.

[3] Shanghai Branch, Hefei National Laboratory, Shanghai, 201315, China.

[4] Shanghai Research Center for Quantum Sciences, Shanghai, 201315, China.
[5] Collaborative Innovation Center of Extreme Optics,
Shanxi University, Taiyuan, 030006, China.

*Corresponding author. Email: guzhibao@sjtu.edu.cn; wpz@sjtu.edu.cn;

**ABSTRACT**

Free induction decay (FID) of spin precession serves as an essential tool for quantum sensing across diverse platforms. While extending spin coherence time remains critical for sensitivity enhancement, the requisite long single-shot acquisitions narrow the resolvable frequency range, establishing a fundamental "spin coherence limit (SCL)", according to the Nyquist Sampling Theorem. Besides, conventional spectral analysis for FID measurement suffers from frequency alias, causing signal attenuation and positional errors that compromise the measurement validity. Here, we demonstrate a general frequency-range-extended technique that, overcomes SCL by leveraging compressive sensing (CS). By applying this method to the FID magnetometer, we expand the resolvable frequency range significantly from the Nyquist-limited range of 251Hz to 3000Hz, effectively avoiding frequency alias. Our work paves the way for applications of high-sensitivity FID measurement in a broad spectral range.

**TEASER**

Our study utilizes compressive sensing to expands the resolvable frequency range of magnetometry without compromising sensitivity.

**INTRODUCTION**

Quantum sensing based on free induction decay (FID) of spin precession has been implemented in a variety of physical systems, including atomic ensembles (*1-6*), nitrogen-vacancy centers in diamond (*7*), nuclear magnetic resonance (*8*), and superconducting quantum interference devices (*9*). Leveraging these systems, FID-based quantum sensor finds extensive applications in biology (*4,5*), imaging (*10*), and fundamental physics (*11*). For precision measurements, improving both sensitivity and measurement frequency range is of critical importance (*4,12-14*). However, spin coherence imposes conflicting constraints on these two parameters, presenting a fundamental challenge in quantum sensing. In particular, quantum mechanics sets a standard quantum limit on the best sensitivity that can be achieved for frequency measurement based on spin precession (*15*):

$$\delta\Omega \simeq \frac{1}{\sqrt{2NFt_\mathrm{m}T}} \tag{1}$$

where $N$ is the number of spins, $F$ is the angular momentum of spins, $T$ is the total measurement time, $t_m$ is the single-shot measurement time. This limit universally constrains the detection of physical fields which can be converted to frequency measurements, including magnetic, electric, and inertial fields, etc. The single-shot measurement time $t_m$ is typically determined by the coherence time $\tau$, which governs the maximum time to reliably accumulate phase information. Consequently, extending the coherence time of quantum systems has long been a central goal for enhancing the performance of spin-based quantum sensors. Remarkable progress has been made in this area, with techniques such as spin locking (*16*), dynamic decoupling (*17*), coherent feedback (*18*) and improved isolation from environmental noise (*19*).

However, while increasing coherence time enhances sensitivity, it simultaneously imposes another limitation on the range of frequencies that can be measured. According to the Nyquist Sampling Theorem (*20*), the maximal measurable frequency range is bounded by:

$$f_{\text{SCL}} = \frac{1}{2\tau}, \qquad (2)$$

which we refer to as the "spin coherence limit (SCL)". Signal components with frequencies exceeding this range are either unresolved or appear as spurious peaks in the detected spectrum. This trade-off between sensitivity and frequency range highlights the fundamental incompatibility between achieving high sensitivity and wide measurement frequency range in precision measurement. Moreover, frequency aliasing poses a critical challenge for applications such as fundamental physics research where distinguishing genuine signals is essential.

In this work, we demonstrate the frequency-range-extended technique based on compressive sensing (CS) to overcome the SCL between sensitivity and frequency range. CS is introduced by Donoho, Candès, Romberg, and Tao two decades ago (*21-26*), and it has been widely used in signal-processing region (*27-39*). CS can overcome the limitations of sampling rates, enabling reconstruction of high-frequency components from sparse signals using sub-Nyquist sampling rates. By applying this technique to the FID magnetometer, the resolvable frequency range is extended from the Nyquist-limited 251 Hz to 3000 Hz while simultaneously suppressing frequency-alias-induced spurious peaks. The modulation signal is reconstructed with the correct amplitude in the right frequency position as well. Our work breaks the longstanding trade-off between sensitivity and frequency range in quantum sensing through CS, opening new possibilities for precision measurements requiring high sensitivity, broad frequency range, and alias-free spectrum.

**RESULTS**

**Measurement Process and Compressive Sensing.**

In this study, to fully utilize coherence time and achieve high sensitivity, the system operates under constrained sampling rates. This leads to undersampling of the $n$-dimensional signal $X$, producing the $m$-dimensional measurement result $Y$ ($m \ll n$). The measurement process is governed by $Y = \Phi X$, where the $m \times n$-dimensional matrix $\Phi$ projects these high-frequency components into a low-dimensional subspace. As shown in Fig. 1, when undersampled with a sampling rate at 7 Hz, the $-5$ Hz and $-6$ Hz components in the original signal $X$ are aliased. As a result, the 1 Hz component in $Y$ consists of both the original 1 Hz component and the frequency component aliased from $-6$ Hz. Hence, when measurements are acquired at a single sampling rate, the system fails to uniquely determine $X$, allowing infinitely many candidate solutions to satisfy $Y = \Phi X$.

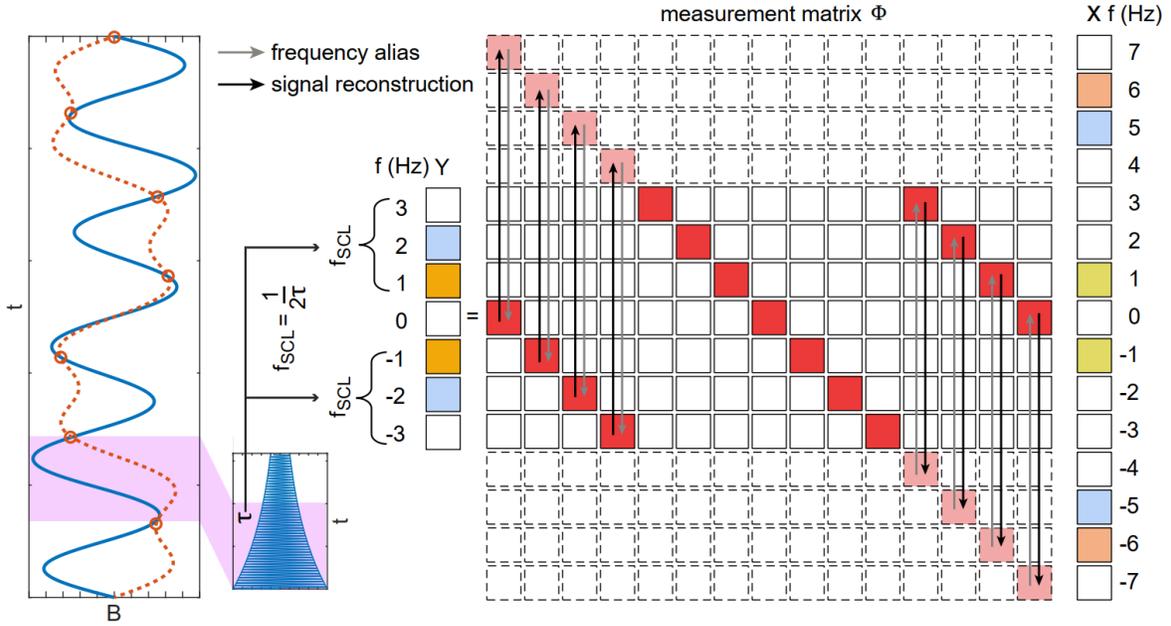

**Fig. 1. The diagram of frequency alias and signal reconstruction.** The original signal (blue line in the left column) contains six spectral components at $\pm 1$ Hz, $\pm 5$ Hz, and $\pm 6$ Hz. When undersampled using the FID measurement, the frequency range of spectrum is limited by coherence time $\tau$, causing aliasing (red dashed line in left column). The resulting spectrum $Y$ exhibits spectral folding. The $-6$ Hz component in the original signal manifests as a $1$ Hz alias in the undersampled data $Y$, while the $-5$ Hz component manifests as a $2$ Hz alias. The CS method aims to reconstruct the signal $X$ with measurement matrix $\Phi$ based on the measurement result $Y$.

Here, we select the multi-rate asynchronous sub-Nyquist (MASS) as the CS scheme (40) to solve the problem. While employing $v$ distinct sampling rates, the joint measurement scheme expand the effective dimensionality of $Y$ and $\Phi$. These measurements must be designed to minimize mutual coherence (i.e., overlap in the information space) to enhance reconstruction fidelity. Meanwhile, when the original signal contains excessive frequency components, aliasing-induced multiple overlaps in the low-frequency domain can corrupt the reconstruction. Thus, CS imposes two requirements: (i) signal sparsity to ensure reconstructability and (ii) measurement incoherence to preserve distinguishable aliasing patterns, enabling robust reconstruction through optimization algorithms. To reconstruct a $k$-sparse spectrum $X$ with length $N_s$, at least $v = 2k - 1$ sub-Nyquist samples $\{Y_1, Y_2, \ldots, Y_v\}$ with corresponding length $\{M_1, M_2, \ldots, M_v\}$ are needed. To ensure that the measurement matrix $\Phi_i$ ($i = 1,2,\ldots v$) satisfies the incoherence property, $M_i$ must be set to be prime numbers of the order of $\sqrt{N_s}$ (40). With respect to $Y_i$, it is sampled at sampling rate $f_i$, and the relation between $Y_i$ and $X$ can be expressed as $Y_i = \Phi_i X$, in which $\Phi_i$ is the $i$-th measurement matrix whose element is $\Phi_i(a,b) = M_i \delta(\mod(\lceil N_s/2 \rceil + \lfloor M_i/2 \rfloor - b, M_i), M_i - a)/N_s$ ($a = 1,2,\ldots m, b = 1,2,\ldots n$) where $\delta$ denotes the Kronecker delta function and "mod" denotes the modulo (remainder) function (see detail in Methods). All sub-Nyquist samples are concatenated vertically as the sampled signal $Y$ and the same operation is performed on all matrices $\Phi_i$ to generate the measurement matrix $\Phi$. By employing Lawson-Hanson algorithm (41), the linear system $Y = \Phi X$ is solved, reconstructing $X$.

After introducing CS, the expanded frequency range is determined by the total measurement time $T$ and coherence time $\tau$. Within $T$, each magnetic field value is acquired over $\tau$. Hence, the length of each under-sampling detection $M_i$ is determined by $M_i \sim T/\tau$. Furthermore, to satisfy CS reconstruction requirements, $M_i$ must be of the order of $\sqrt{N_s}$ (i.e., $M_i \sim \sqrt{N_s}$). Here, $N_s$ is the length of reconstructed spectrum $X$, which relates to the CS-enhanced frequency range $f_{CS}$ and frequency resolution $1/T$ through $N_s = 2Tf_{CS}$. Considering these relations mentioned above, the frequency range should be constrained as:

$$f_{CS} \simeq \frac{T}{2\tau^2} = \frac{T}{\tau}f_{SCL}. \tag{3}$$

Fig. 2 shows trade-off relations between normalized sensitivity of Larmor frequency $\delta\Omega_0 = \sqrt{2NF}\,\delta\Omega = 1/(\tau T)$ and frequency range with and without CS. Blue and red lines denote the SCL and the CS-enhanced limit, respectively. $T$ is set to be 1 s and 10 s for solid lines and dashed lines, respectively. An extended $T$ allows for a greater number of sampling points under a fixed $\tau$, thereby showing a better spectral extension performance via CS.

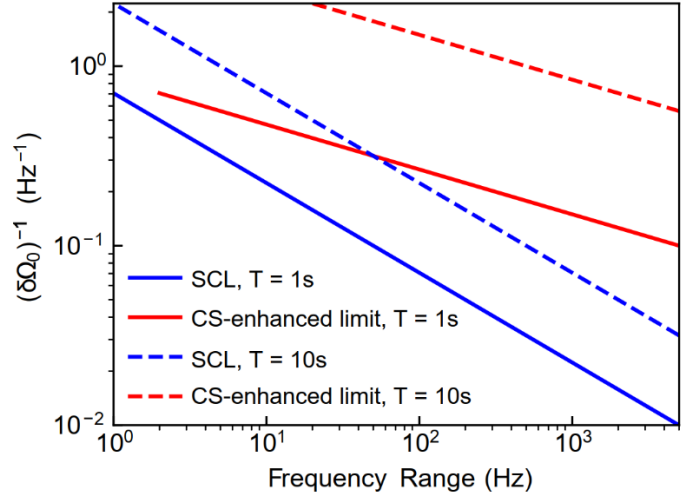

**Fig. 2. The trade-off relations between normalized sensitivity of Larmor frequency $\delta\Omega_0$ and frequency range.** The blue lines represent the SCL, while the red lines represent the CS-enhanced limit. Solid and dashed lines correspond to $T = 1\,s$ and $T = 10\,s$, respectively. By employing CS, the frequency range is extended while maintaining the same sensitivity.

**Experimental Setup**

The experimental setup is illustrated in Fig. 3A. The cylindrical atomic cell containing isotopically enriched $^{87}$Rb atoms is placed within a five-layer magnetic shield. The leading static magnetic field $B_0$ is set as 3.8 μT. Three lasers are employed in this experiment as pump, repump, and probe. The pump laser is resonant with the $^{87}$Rb D1 line $F = 2$ to $F' = 2$ transition, polarizing the atoms along the $-\hat{y}$ direction. The repump laser is resonant with the $^{87}$Rb D1 line $F = 1$ to $F' = 2$ transition, exciting atoms from the $F = 1$ ground state to enhance atoms' spin polarization. The probe laser is 3 GHz red detunning from the $^{87}$Rb D2 $F = 2$ to $F' = 2$ transition. The pump and repump lasers combined and adjusted as rectangular pulses with 3% duty cycle ratio at the Larmor frequency 26.6 kHz with an AOM. They are transformed into left-circularly polarized using a quarter-wave plate and then pass through the atomic cell along the $-\hat{y}$ direction. The powers of pump and repump lasers are 2 μW and 100 μW with 3% duty cycle ratio, respectively. The power of probe laser is 200 μW.

In a single pump-probe cycle, after synchronized pumping and repumping for time duration $t_{\text{pump}}$, atoms are fully polarized. The pump and repump lasers are then turned off, allowing the atomic spins to process freely in the $\hat{x} - \hat{y}$ plane with Larmor frequency for a time duration $t_{\text{probe}}$. The $\hat{y}$-polarized probe laser goes through the cell in $\hat{x}$ direction and experiences polarization rotation after passing through the atomic cell. The rotation angle is proportional to the atom polarization along $\hat{x}$ direction. The output optical polarization rotation is measured with a balanced polarimeter upon transmission through the cell, as FID signal. The FID signal takes the form of a sine wave with exponential decay (6)

$$S(t) = A_0 \cdot e^{-\frac{t}{\tau}} \cdot \sin(2\pi\Omega t + \phi_0), \qquad (4)$$

in which $A_0$ is the maximum amplitude, $\tau$ denotes the coherence time, $\Omega = \gamma_B B$ is the Larmor frequency, $\gamma_B$ is the gyromagnetic ratio, $\phi_0$ is the initial phase, and $t$ is the time variable. The magnetic field $B$ is extracted from each pump-probe cycle by fitting the FID signal over the time duration $t_{\text{probe}}$. The sensitivity of $B$ is optimized when $t_{\text{probe}} = \tau$. Consequently, the sampling rate is limited to $1/(t_{\text{pump}} + \tau)$, which also restricts the resolvable frequency range.

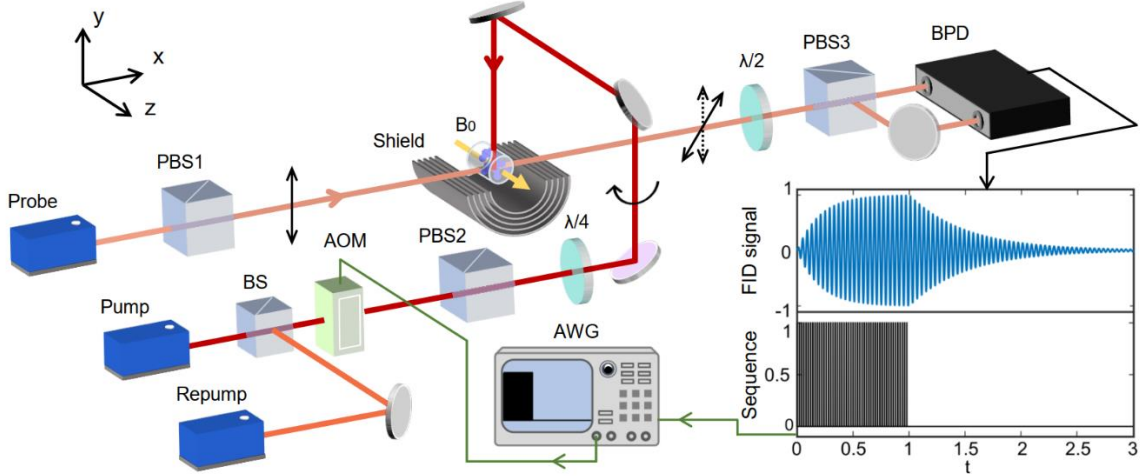

**Fig. 3. Experimental setup.** PBS: polarizing beam splitter. BS: beam splitter. AOM: acousto-optic modulator used to pulse the pump and repump laser. AWG: arbitrary waveform generator. $\lambda/2$, half-wave plate. $\lambda/4$: quarter-wave plate. BPD: balanced photodetector. Atoms are contained in a vapor cell positioned in the center of the magnetic shield and are pumped and probed by laser beams under a static magnetic field $B_0$ (along $\hat{z}$). The insert shows the FID signal and the timing control pulse sequences of the pump and repump lasers.

**Experimental Results.**

To examine the capability of CS in extending the resolvable frequency range, an oscillating magnetic field with a frequency of $800\,\text{Hz}$ (above the SCL) and a magnitude of $1.17\,\text{nT}$ (as shown in Fig. 4A) is applied by a pair of Helmholtz coils along $\hat{z}$-direction. We acquire 41 sets of data with the prime sampling rates ($\nu = 41$) from $263\,\text{Hz}$ to $503\,\text{Hz}$. The total sampling duration $T$ was consistently maintained at $1\,\text{s}$ for each measurement, while the sampling rates $M_\nu$ are controlled by adjusting the $t_{\text{probe}}$.

The recorded time-domain signals sampled at these rates are illustrated in Fig. 4(B, D, F). To extract the AC magnetic field characteristics, all signals are analyzed using discrete Fourier transform (DFT). The resulting amplitude spectra are presented in Fig. 4(C, E, G). Due to under-sampling scenarios, the inevitable frequency alias and spectral overlap distort the authenticity of the spectrum. Even though those spectra originate from the same applied magnetic field, they are all different from each other and unreliable. For instance, the three presented spectra in Fig. 4(C, E, G) display dominant peaks at $12\,\text{Hz}$, $43\,\text{Hz}$, and $207\,\text{Hz}$, respectively, all of which are frequency aliases of the applied signal at $800\,\text{Hz}$. This means that even those frequency components within the bandwidth could be misrepresented in under-sampled spectra.

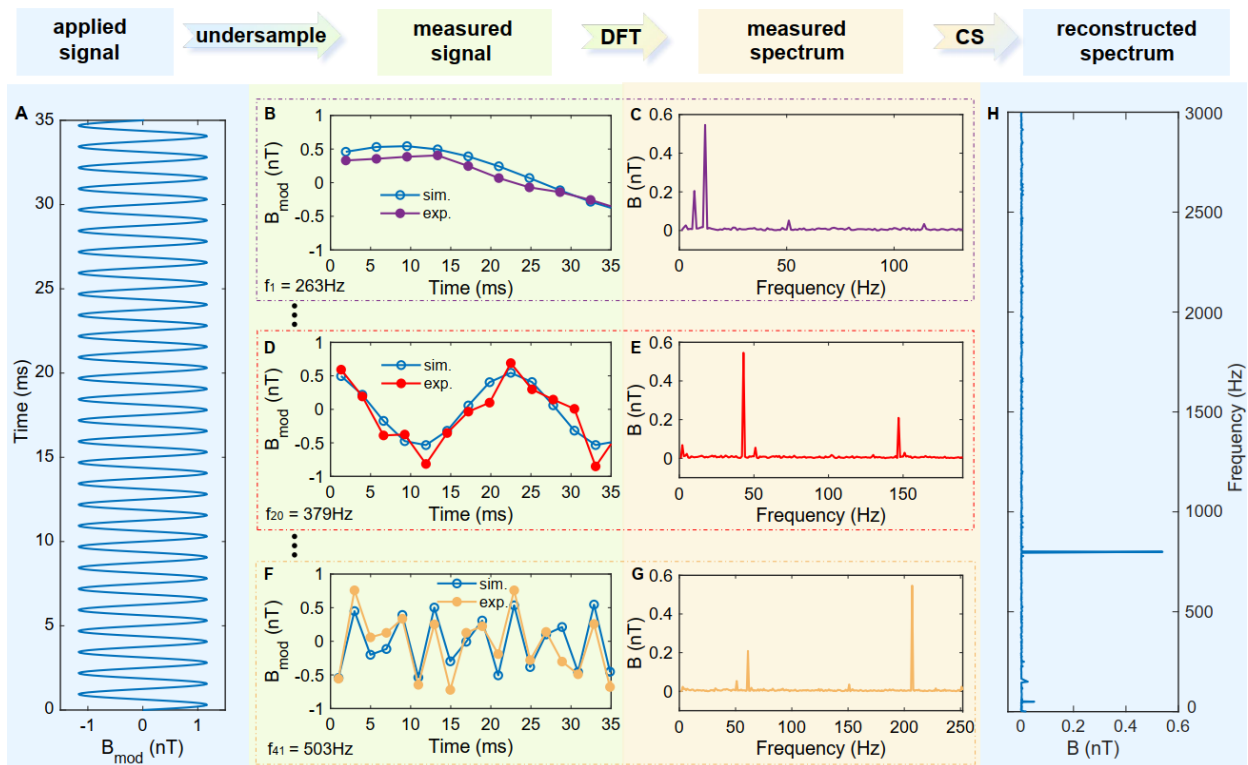

**Fig. 4. Wideband spectral reconstruction based on CS.** (**A**) The applied magnetic field modulation with frequency of 800 Hz. (**B**), (**C**), and (**F**) are magnetic field variations over time obtained using undersampled rates with selected repetition rates at 263 Hz, 379 Hz, and 503 Hz, where open circles represent theoretical simulation results and solid circles denote experimental measurements. (**C**), (**E**), and (**G**) are the corresponding undersampled spectra derived from the time-domain magnetic field signals. (**H**) Reconstructed wideband spectrum containing the 800 Hz modulation signal.

These spectra are vertically concatenated to form the final measurement result $Y$. Together with the measurement matrix $\Phi$, the real spectrum $X$ is reconstructed. Fig. 4H shows the reconstructed noise spectrum of the magnetic field, which has an extended frequency range to 3000 Hz. The reconstructed spectrum reveals four dominant frequency components: 50 Hz and its harmonics at 150 Hz and 250 Hz (induced by AC power line), along with the applied 800 Hz modulation signal. The sparsity $k$ of the reconstructed bilateral spectrum is 8. Since the number of measurements $v = 41$ satisfies $v > 2k - 1$, the condition required by MASS scheme is fulfilled, ensuring accurate reconstruction of the sparse spectrum. Although the frequency range is extended, the nature of phase accumulation in the measurement process still induce a low-pass filtering effect on magnetic field. The correspond frequency response is provided in the Methods. After correcting the measured 800 Hz amplitude in Fig. 4H using the frequency response, the adjusted value matches the applied magnetic field intensity in Fig. 4A. The consistency between reconstructed and applied magnetic field amplitudes is shown in Methods with an error of 0.6%.

The sensitivity of magnetometry with (orange line) and without CS (blue line) are compared in Fig. 5. Here, pump-probe repetition rate is set to 503 Hz. For conventional FID magnetometry, two aliasing peaks appear at 60 Hz and 206 Hz, with its resolvable frequency range limited to half of the pump-probe repetition rate, which here is 251 Hz. In contrast, the CS-enhanced implementation eliminates spectral alias while extending the frequency range to 3000 Hz.

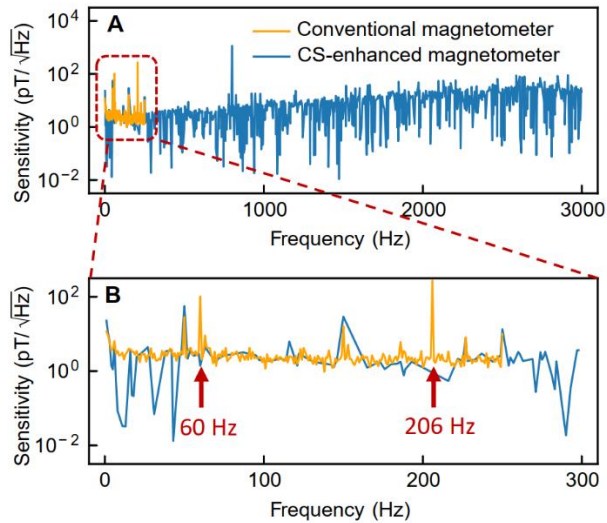

**Fig. 5. Sensitivity Comparison.** (**A**) Sensitivity of the conventional FID magnetometer and the CS-enhanced FID magnetometer. (**B**) An enlarged view of the area enclosed by the dashed line in (A). The sensitivity of both magnetometer configurations remains consistent, with PSN limited sensitivity of $4\,\mathrm{pT}/\sqrt{\mathrm{Hz}}$ at $100\,\mathrm{Hz}$. Due to the presence of high-frequency signals exceeding the Nyquist frequency, the conventional magnetometer exhibits significant spurious peaks at $60\,\mathrm{Hz}$ and $206\,\mathrm{Hz}$ (marked by red arrows).

## DISCUSSION

We address the fundamental challenge in quantum sensing: the trade-off between sensitivity and measurement frequency range. While conventional FID magnetometers are limited by the Nyquist frequency and suffer from aliasing when measuring high-frequency components, we demonstrate a novel approach using CS to expand the frequency range of the FID magnetometer by a factor of 12, achieving a frequency range of 3000 Hz.

In this work, the sensitivity in the extended frequency range is governed by photon shot noise (PSN). Considering the low-pass filtering effect of magnetic signal, the sensitivity exhibits a characteristic degradation with increasing frequency (*42*). When limited by spin projection noise, which also decreases with frequency following a Lorentzian trend, the magnetometer's sensitivity remains frequency-independent (*43*).

The advancement of developed approach with high sensitivity, broad frequency range, and alias-free spectrum is particularly beneficial for quantum systems with long coherence times, such as atomic spin (*44*), nuclear spins (*18,45*), nitrogen-vacancy centers (*7,46*), Bose-Einstein condensates (*47*), and rare-earth elements (*19*), offering a valuable tool for precision measurements in quantum sensing applications.

## MATERIALS AND METHODS
### Frequency response

In order to obtain the frequency response of the magnetometer, we apply a sinusoidal modulated magnetic field along $\hat{z}$ direction with amplitude of $1.14\,\mathrm{nT}$. The magnetic modulation appears as a peak in the spectrum of the magnetometer signal and its amplitude decreases as a low-pass filter with the increase of the magnetic modulation frequency as shown in Fig. 6.

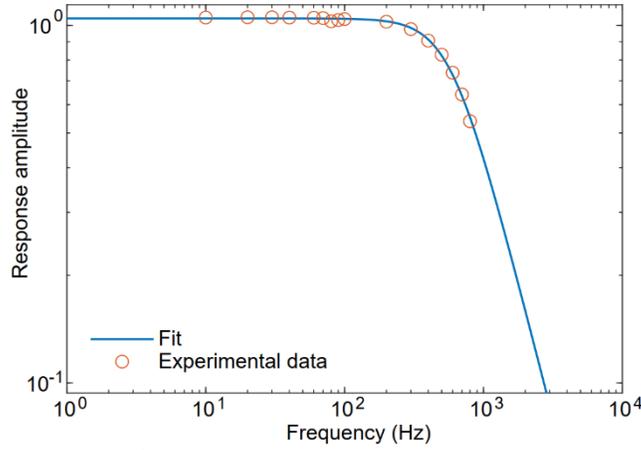

**Fig. 6. The frequency response of the magnetometer.** The circles show the peak of magnetic signal with sinusoidal modulated magnetic field when varying the magnetic modulation frequency and the solid line shows fit with the transfer function of low-pass filter.

**Reconstruction consistency**

Theoretically, there should be a linear relationship with a slope of 1 between our reconstructed amplitude $B_{rec}$ and the applied amplitude of the magnetic field modulation $B_{mod}$. To demonstrate this, we preset multiple groups of magnetic field modulation at a frequency of 137 Hz. We then used our experimental system to measure these fields and obtained the magnetic field values by CS. The results are shown in Fig. 7. The red circles represent our experimental data points, and the blue solid line is the linear fit function with a slope of 1.006. Hence, we can conclude that our experimental system can reflect the actual amplitude of the magnetic field modulation.

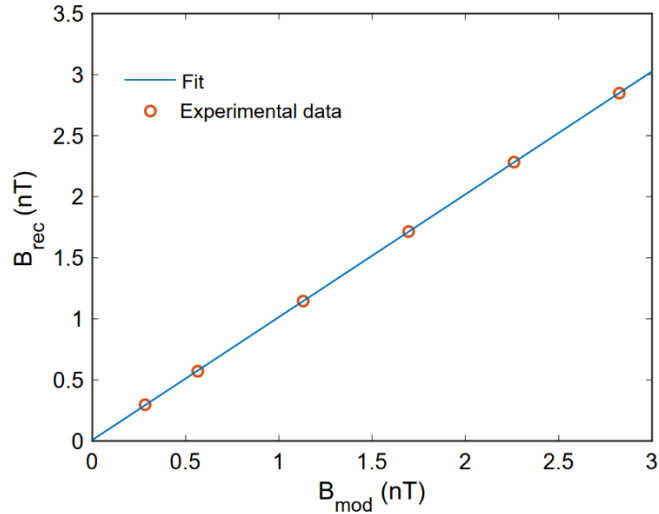

**Fig. 7.** The relationship of the reconstructed magnetic field amplitude $B_{rec}$ and the applied magnetic modulation amplitude $B_{mod}$. The circles are experimental data and the solid line is the fit with linear function.

**Construction of the measurement matrix**

The DFT process between spectrum $Y[m]$ and discrete sampling data $x[t]$ as sampling rate $f_i$ is:

$$Y[m] = \sum_{k=0}^{N_i-1} x[t] e^{\frac{-2\pi i t m}{N_i}} \qquad (5)$$

where $N_i$ is the length of samples. For $m \in [0, \lfloor N_i/2 \rfloor]$, $Y[m]$ denotes the component with frequency $f = mf_i/N_i$, while $m \in [\lceil N_i/2 \rceil, N-1]$, $Y[m]$ denotes the component with frequency $f = -(N_i - m)f_i/N_i$. There's a mapping relation $F_i(m)$ that maps the index $m$ towards the corresponding frequency component:

$$F_i(m) = \begin{cases} m \cdot \dfrac{f_i}{N_i} & m \in \left[0, \left\lfloor \dfrac{N_i - 1}{2} \right\rfloor\right] \\ (m - N_i) \cdot \dfrac{f_i}{N_i} & m \in \left[\left\lceil \dfrac{N_i + 1}{2} \right\rceil, N_i - 1\right] \end{cases}, \quad (6)$$

and its inverse mapping relation is:

$$F_i^{-1}(f) = \begin{cases} \dfrac{f \cdot N_i}{f_i} & f \in \left[0, \left(\left\lceil \dfrac{N_i + 1}{2} \right\rceil - 1\right) \cdot \dfrac{f_i}{N_i}\right] \\ \dfrac{f \cdot N_i}{f_i} + N_i & f \in \left[\left(\left\lceil \dfrac{N_i + 1}{2} \right\rceil - N_i\right) \cdot \dfrac{f_i}{N_i}, 0\right] \end{cases}, \quad (7)$$

which maps the frequency $f$ towards the index. The continues signal $x(t)$ has its spectrum $x_c(f)$ as $x_c(f) = \int x(t)e^{-i2\pi f \cdot t} dt$. Denoting the bandwidth of spectrum $x_c(f)$ as $f_{BW}$, for a DFT spectrum obtained with sampling rate $f_i < 2f_{BW}$, the component with frequency $f$ is a summation over $X_c(f)$:

$$Y[F_i^{-1}(f)] = f_i \sum_{l=-\infty}^{\infty} x_c(f + lf_i). \quad (8)$$

While the signal is sampled with a frequency $f_s > 2f_{BW}$, the discrete spectrum without frequency alias, $X[F_s^{-1}(f)]$ is proportional to $x_c(f)$:

$$X[F_s^{-1}(f)] = f_s x_c(f). \quad (9)$$

According to the upper two relations, the equation

$$Y[F_i^{-1}(f)] = \frac{f_i}{f_s} \sum_{l=l_{min}}^{l_{max}} X[F_s^{-1}(f)] \quad (10)$$

holds when $f + lf_i \in [-f_s/2, f_s/2]$. The summation range for $l$ is limited to:

$$l_{min} = \left\lceil \left\lceil \frac{N_s}{2} \right\rceil \frac{1}{N_i} - \frac{N_s}{N_i} - \frac{F_i(m)N_s}{N_i f_s} \right\rceil, l_{max} = \left\lfloor \left\lceil \frac{N_s}{2} \right\rceil \frac{1}{N_i} - \frac{1}{N_i} - \frac{F_i(m)N_s}{N_i f_s} \right\rfloor. \quad (11)$$

The upper equation could be noted as $Y_i = \Theta_i X_i$ in which $\Theta_i$ represents an $m \times n$ matrix whose $(a+1, b+1)$ entry is:

$$\Theta_i(a+1, b+1) = \frac{f_i}{f_s} \sum_{l=l_{min}}^{l_{max}} \delta\left(b, F_s^{-1}(F_i(a) + lf_i)\right) \quad (12)$$

To better highlight the two-sided spectral features of $X$ and $Y$, one can apply the following linear transformations to each: $X' = U_s X, Y' = U_i Y$. Here, $U_s$ and $U_i$ each have the following block-diagonal form:

$$U_s = \begin{pmatrix} 0 & I'_s \\ I_s & 0 \end{pmatrix}, U_i = \begin{pmatrix} 0 & I'_i \\ I_i & 0 \end{pmatrix} \quad (13)$$

$I'_s$ and $I_s$ are the $\lceil N_s/2 \rceil$-dimensional and $\lfloor N_s/2 \rfloor$-dimensional identity matrix, respectively. $I'_i$ and $I_i$ denotes identity matrix with dimension $\lceil N_i/2 \rceil$ and $\lfloor N_i/2 \rfloor$, respectively. Let the measurement matrix after the above transformations be $\Phi$, then the relationship between $\Phi_i$ and $\Theta_i$ can be written as $\Phi_i = U_i \Theta_i U_s^{-1}$. The element of $\Phi_i(a,b)$ is:

$$\Phi_i(a,b) = \frac{N_i}{N_s} \delta \left( Mod \left( \left\lceil \frac{N_s}{2} \right\rceil + \left\lfloor \frac{M_i}{2} \right\rfloor - b, M_i \right), M_i - a \right) \qquad (14)$$

**ACKNOWLEDGMENTS**
This work is supported by the Innovation Program for Quantum Science and Technology (2021ZD0303200); the National Natural Science Foundation of China (12234014, 12204303, 11654005); the Fundamental Research Funds for the Central Universities; the Shanghai Municipal Science and Technology Major Project (2019SHZDZX01); Shanghai Science and Technology Innovation Action Plan (24LZ1401400); Innovation Program of Shanghai Municipal Education

Commission (202101070008E00099); the National Key Research and Development Program of China (Grant No. 2016YFA0302001); and the Fellowship of China Postdoctoral Science Foundation (Grant No. 2021M702150); W.Z. also acknowledges additional support from the Shanghai talent program.

**AUTHOR CONTRIBUTIONS:** W.Z. supervised the whole project. G.B. and W.Z. conceived the research. G.B., and W.Z. designed the experiments. G.B., R.W., P.Y., and D.H. performed the experiment. G.B. and R.W. contributed to the theoretical study. G.B., R.W., and P.Y. analyzed the data. G.B., R.W., and P.Y. draw the diagrams. G.B., R.W., P.Y., and W.Z. wrote the paper. All authors contributed to the discussion and review of the manuscript.